\documentclass[11pt,letterpaper]{article}
\usepackage{dblfloatfix}
\usepackage{bbm}
\usepackage{blindtext}
\usepackage{algpseudocode}
\usepackage{amsmath}
\usepackage{amssymb}
\usepackage{amsthm}
\usepackage{color}
\usepackage{cite}
\usepackage{graphicx}
\graphicspath{{./}{Figs/}}
\usepackage{subcaption}
\usepackage{mathrsfs}
\usepackage{verbatim}
\usepackage{indentfirst}
\usepackage[titletoc,toc,title]{appendix}
\usepackage{appendix}
\usepackage{url}
\usepackage{epsfig,endnotes,amsthm,changepage}
\usepackage{authblk}

\usepackage{epstopdf,lipsum,etoolbox}

\usepackage{graphicx}
\usepackage{amsfonts}
\usepackage{setspace}
\usepackage{cite}
\usepackage{array}
\usepackage{mdwmath}
\usepackage{mdwtab}
\usepackage{mdwtab, stmaryrd}
\usepackage{times}
\usepackage{epsfig}
\usepackage{latexsym}
\usepackage{epstopdf}
\usepackage{verbatim}
\usepackage{units}
\usepackage{amsthm}
\usepackage{mdwlist}
\usepackage{dblfloatfix}
\usepackage{blindtext}
\usepackage{array}
\newcolumntype{P}[1]{>{\centering\arraybackslash}p{#1}}
\usepackage{tabu}

\AppendGraphicsExtensions{.pdf}
\usepackage{epstopdf,lipsum,etoolbox}

\newtheorem{theorem}{Theorem}

\newtheorem*{proposition*}{Proposition}

\allowdisplaybreaks

\begin{document}

\title{Communication Efficient Federated Learning over Multiple Access Channels}

 \author{{Wei-Ting~Chang \qquad Ravi~Tandon}\\
 Department of Electrical and Computer Engineering\\
 University of Arizona\\
 Email: \{\textit{wchang, tandonr}\}@email.arizona.edu
 }

\maketitle

\begin{abstract}
In this work, we study the problem of federated learning (FL), where distributed users aim to jointly train a machine learning model with the help of a parameter server (PS). In each iteration of FL, users compute local gradients, followed by transmission of the quantized gradients for subsequent aggregation and model updates at PS. One of the challenges of FL is that of communication overhead due to FL's iterative nature and large model sizes. One recent direction to alleviate communication bottleneck in FL is to let users communicate simultaneously over a multiple access channel (MAC), possibly making better use of the communication resources. 

In this paper, we consider the problem of FL learning over a MAC. In particular, we focus on the design of digital gradient transmission schemes over a MAC, where gradients at each user are first quantized, and then transmitted over a MAC to be decoded individually at the PS. When designing digital FL schemes over MACs, there are new opportunities to assign different amount of resources (such as rate or bandwidth) to different users based on a) the informativeness of the gradients at each user, and b) the underlying channel conditions. We  propose a stochastic gradient quantization scheme, where the quantization parameters are optimized based on the capacity region of the MAC. We show that such \textit{channel aware quantization} for FL outperforms uniform quantization, particularly when users experience different channel conditions, and when have gradients with varying levels of informativeness.


\end{abstract}


\section{Introduction}
\footnote{This work was supported by US NSF through grants CAREER 1651492, CNS 1715947, and by the Keysight Early Career Professor Award.}
Federated Learning (FL) refers to a distributed machine learning (ML) framework that allows distributed machines, or users, to collaboratively train an ML model with the help of a parameter server (PS). Typically, users compute gradients for a global model on their local data, and send gradients to the PS for aggregation and model updates in an iterative fashion. FL is appealing and has gained recent attention due to the fact that it allows natural parallelization, and can be more efficient than centralized approaches in terms of storage. However, communication overhead caused by exchanging gradients remains an issue that needs to be addressed.

Previous works alleviate the communication bottleneck by compressing gradients before transmissions. Two commonly used gradient compression approaches are $a)$ quantization, and $b)$ sparsification. Gradient quantization follows the idea of lossy compression by describing gradients using a small number of bits and these low-precision gradients are transmitted back to the PS. One extreme is to send just $1$ bit of information per value \cite{1BitSGD}. Similar idea was used in signSGD \cite{signSGD2018} and TernGrad \cite{TernGrad2017}, which use $1$ and $2$ bits to describe each value, respectively. In gradient sparsification, some coordinates of the gradient vector are dropped based on certain criteria \cite{sparse-1, sparse-2}, which for instance, can depend on the variance and informativeness of the gradients. Other quantization/sparsification techniques include \cite{sparse-3,sparse-4,sparse-5,QSGD,klevel}. However, these stand alone compression techniques are not tuned to the underlying communication channel over which the exchange takes place between the users and the PS, and may not utilize the channel resources to the fullest.

Another line of recent works study FL over wireless channels, and more generally multiple access channels (MACs).  The superposition nature of wireless channels allows gradients to be aggregated "over-the-air" and allows for much more efficient training. Several recent works include \cite{AmiDen2019,AmiDen2019-2,AmiDen2019-3,AmiDum2019,AbaOzf2019,CheYan2019,YanJia2018,ZenDu2019,ZhuWan2020,SunZho2019,WanTuo2018,SerCoh2019,SerCoh2019-2}. The approaches can be broadly categorized into digital or analog schemes depending on how the gradients are transmitted over the channel. In analog schemes, the local gradients are scaled and directly transmitted over the wireless channel, allowing PS to directly receive a noisy version of the aggregated gradient. In digital schemes, gradients from users are decoded individually, but transmission still occurs over a MAC.  
Although it has been shown that in terms of bandwidth efficiency, analog schemes can be superior than digital schemes \cite{AmiDen2019,AmiDen2019-3}, we argue that digital schemes  have the following advantages: $a)$ backward compatibility - they can be easily implemented on the existing digital systems, $b)$ they are less prone to slow users, $c)$ they are more reliable due to the fact that various error control codes can be used, and $d)$ digital schemes do not require tight synchronization as required by analog transmission.

\textit{Main Contributions:} Motivated by the above discussion, we consider FL learning over a MAC and focus on the design of digital gradient transmission schemes, where gradients at each user are first quantized, and then transmitted over a MAC to be decoded individually at the PS. When designing digital FL schemes over MACs, we show that there are new opportunities to assign different amount of resources (such as rate or bandwidth) to different users based on a) the informativeness of the gradients at each user, and b) the underlying channel conditions. We  propose a stochastic gradient quantization scheme, where the quantization parameters are optimized based on the capacity region of the MAC. We show that such \textit{channel aware quantization} for FL outperforms channel unaware quantization schemes (such as uniform allocation), particularly when users experience different channel conditions, and when have gradients with varying levels of informativeness.


\section{System Model \label{Sec:SystemModel}}
\begin{figure}[t]
\centering
	\includegraphics[width=0.55\linewidth]{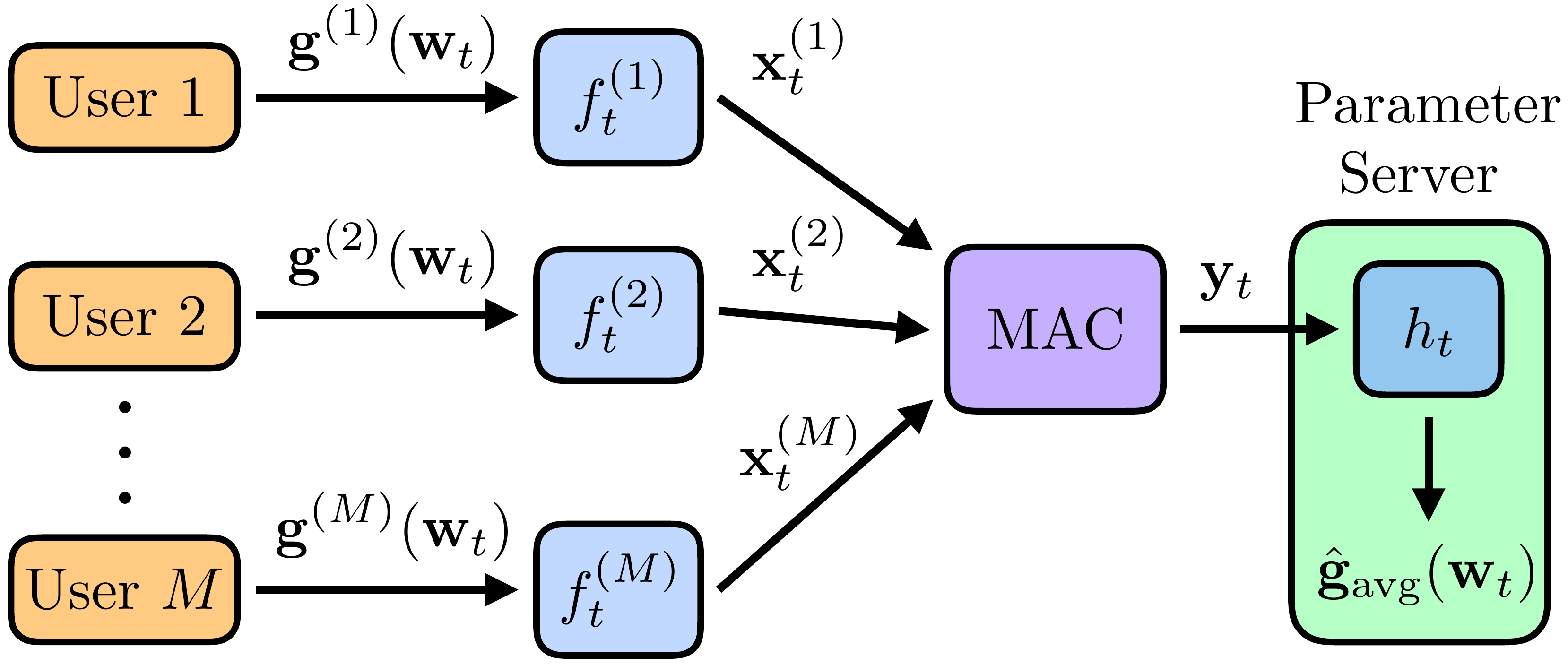}
	\caption{FL over a MAC. At each iteration, users send local gradients $\mathbf{g}^{(m)}(\mathbf{w}_t)$ through a MAC. The PS aggregates the gradients, updates the model and sends the updated model to users for subsequent iteration.
	\label{Fig:model}}
	\vspace{-15pt}
\end{figure}
We consider a distributed machine learning system with a parameter server (PS) and $M$ users, where users are connected to the PS through a Gaussian MAC as shown in Fig. \ref{Fig:model}. Users want to collaboratively train a machine learning model $\mathbf{w}$ with the help of PS by minimizing an empirical loss function,
\begin{align}
L(w) = \frac{1}{M}\sum\limits_{m=1}^{M}\frac{1}{n_m}\sum\limits_{d_n^{(m)}\in\mathcal{D}^{(m)}}\ell(\mathbf{w}, \mathbf{d}_n^{(m)}),
\end{align}
where $\mathcal{D}^{(m)},~|\mathcal{D}^{(m)}|=n_m,~m=1,\dots,M$ denotes the local data set at user $m$ and $\mathbf{d}_n^{(m)}$ is the $n$-th data point in $\mathcal{D}^{(m)}$, and $\ell(\cdot)$ is the loss function. The minimization is done by using gradient descent (GD) algorithm. Each user computes the local gradient $\mathbf{g}^{(m)}(\mathbf{w}_t)\in\mathbb{R}^{d}$ on the local data set $\mathcal{D}^{(m)}$, where $\mathbf{w}_t$ is vector of model parameters at iteration $t$, and
\begin{align}
    \mathbf{g}^{(m)}(\mathbf{w}_t) = \frac{1}{n_m}\sum\limits_{n=1}^{n_m} \triangledown \ell(\mathbf{w}_t, \mathbf{d}_n^{(m)}),~\mathbf{d}_n^{(m)}\in \mathcal{D}^{(m)},~\forall m.
\end{align}
At each iteration, each user $m$ sends a function of its computed gradient $\mathbf{x}_{t}^{(m)}=f_{t}^{(m)}(\mathbf{g}^{(m)}(\mathbf{w}_t))$ back to the PS through $s$ channel uses of the MAC, where $f_{t}^{(m)}(\cdot)$ is some pre-processing function the PS assigned to user $m$ at iteration $t$. We note that the capacity region of a Gaussian MAC can be described as follows \cite{Cover-thomas},
\begin{align}
    \sum\limits_{m\in\mathcal{M}} r_m \leq C_{\mathcal{M}},~\mathcal{M}\subset [M],~|\mathcal{M}|=1,\dots,M,
\end{align}
where $r_m$ denotes the transmission rate of user $m$ and $C_{\mathcal{M}}$ denotes the sum capacity of the users in subset $\mathcal{M}$. We assume an average transmit power constraint $P_m$ for user $m$, and in this case, $C_{\mathcal{M}} = 0.5\log(1+ \sum_{m\in \mathcal{M}} P_m/\sigma^{2})$, where $\sigma^{2}$ denotes variance of the channel noise.

At iteration $t$, the received signal at the PS $\mathbf{y}_t$ is a function of all $\mathbf{x}_t^{(m)}$. The goal of the PS is to recover the average of the local gradients $\mathbf{g}_{\text{avg}}(\mathbf{w}_t)=\sum_{m=1}^{M}\mathbf{g}^{(m)}(\mathbf{w}_t)/M$ from $\mathbf{y}_t$ using some post-processing function $h_t(\cdot)$. However, due to the pre- and post-processing, and the capacity region of the MAC, the PS can only recover the noisy versions of the local gradients $\hat{\mathbf{g}}^{(m)}(\mathbf{w}_t)$, thus, the noisy version of the average gradient $h_t(\mathbf{y}_t)=\hat{\mathbf{g}}_{\text{avg}}(\mathbf{w}_t)=\sum_{m=1}^{M}\hat{\mathbf{g}}^{(m)}(\mathbf{w}_t)/M$. Therefore, the transmission from the users must ensure that the gradients received at the PS are unbiased estimators of $\mathbf{g}^{(m)}(\mathbf{w}_t)$ and have bounded variance, i.e.,
\begin{align}
\mathbb{E}\left[ \hat{\mathbf{g}}^{(m)}(\mathbf{w}_t) \right] = \mathbf{g}^{(m)}(\mathbf{w}_t),~ \text{Var}(\hat{\mathbf{g}}^{(m)}(\mathbf{w}_t)) \leq \epsilon_m,
\end{align}
where the variance bound $\epsilon_m$ should be as small as possible. 

\noindent \textbf{Problem Statement}
When jointly transmitting over a MAC, it is critical to allocate resources efficiently to ensure that the gradient aggregation can be done in a timely manner, and the training error is low. Let $\{r_1,\dots,r_M\}$ be the set of rates allocated to users for gradient transmission over the MAC. In this work, we want to understand how one should allocate rates as a function of the capacity region of the MAC, and the underlying informativeness of the gradients at different users. Furthermore, we want to characterize the resulting trade-off between the underlying channel conditions of the MAC and the convergence rate of GD algorithms.

\section{Main Results}
In this section, we present our proposed stochastic gradient quantization scheme for GD, which is inspired by schemes in \cite{klevel,AgaSur2018}. In this scheme, the PS asks users to quantize their local gradients before sending them based on individual quantization budgets. The quantization budgets are found by the PS by solving an optimization problem that aims to minimize the variance of the aggregated gradients,  while satisfying the transmission rate constraints imposed by the MAC. The distinction between our scheme and the scheme in \cite{klevel} is that we allow each user to have its own quantization budget. We first present the proposed scheme for any number of users $M$, analyze the convergence rate of the scheme, and present a general optimization problem for quantization budget allocation based on the capacity of the MAC. We then show an example with $M=2$ users and solve for the optimal quantization budgets and communication rates.

\subsection{Stochastic Multi-level Gradient Quantization}
\begin{figure}[t]
\centering
	\includegraphics[width=0.5\linewidth]{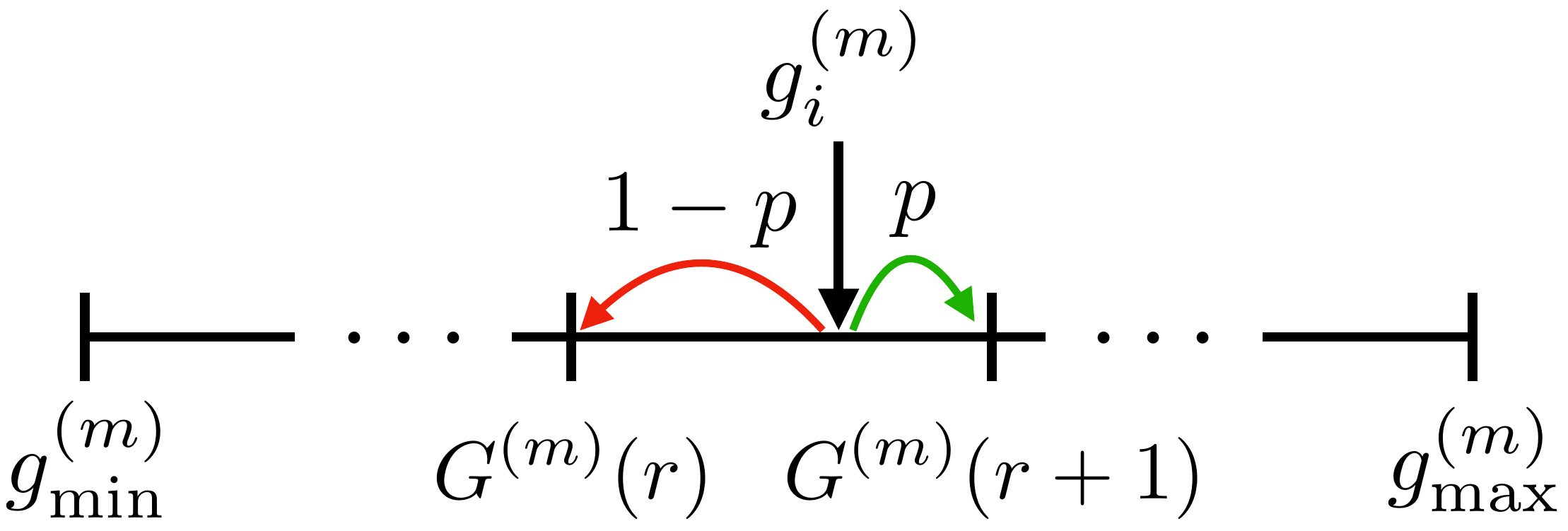}
	\caption{Stochastic multi-level gradient quantization where the dynamic range of the gradient vector is split into $k_m$ levels. Subsequently, each element of the vector $\mathbf{g}_i^{(m)}$ is quantized to $G^{(m)}(r)$ with probability $p$ as shown in \eqref{eq:Qg}, or to $G^{(m)}(r+1)$ with probability $1-p$.
	\label{Fig:multilevel}}
	\vspace{-15pt}
\end{figure}

At each iteration $t$, each user $m$ computes the local gradient vector $\mathbf{g}^{(m)}(\mathbf{w}_t)$ using its local data set $\mathcal{D}_{t}^{(m)},~m=1,\dots,M$. For simplicity of notation, we drop the iteration index $t$ in describing the quantization scheme. Each user computes the dynamic range of its local gradient, i.e., $\Delta_m=g_{\max}^{(m)}-g_{\min}^{(m)}$, where $g_{\max}^{(m)}$ and $g_{\min}^{(m)}$ are the maximum and minimum values of the local gradient vector at user $m$. The user then quantizes its local gradient vector using the stochastic multi-level quantization scheme as we describe next. For every integer $r\in [0,k_m)$, we define
\begin{align}
G^{(m)}(r) \triangleq g_{\min}^{(m)} + \frac{r\Delta_m}{k_m-1},\label{eq: klevelDef}
\end{align}
where $k_m\geq 2$ is the quantization budget for user $m$. For each element $i$ in the local gradient vector, if $g_i^{(m)}\in [G^{(m)}(r),G^{(m)}(r+1))$, then $g_i^{(m)}$ is quantized as follows,
\begin{align}
Q\big(g^{(m)}_{i}\big) =
  \begin{cases}
    G^{(m)}(r+1)     &  \text{w.p. } \frac{g^{(m)}_{i} - G^{(m)}(r)}{ G^{(m)}(r+1) -G^{(m)}(r)} \\
    G^{(m)}(r)   &  \text{otherwise}
  \end{cases}.\label{eq:Qg}
\end{align}
This operation is shown in Fig. \ref{Fig:multilevel}. Once the entire gradient vector is quantized, user $m$ sends its quantized gradient vector $\mathbf{Q}(\mathbf{g}^{(m)})= [Q\big(g^{(m)}_{1}\big), \ldots, Q\big(g^{(m)}_{d}\big)]$ to the PS over the Gaussian MAC. We assume that before each iteration, each user describes the scalars $g_{\max}^{(m)}$ and $g_{\min}^{(m)}$ (which describe the dynamic range $\Delta_m = g_{\max}^{(m)} - g_{\min}^{(m)}$ of the local gradient) at full resolution to the PS. In addition, as each element in the gradient vector is quantized to be one of the $k_m$ levels, hence, a total of $d\log_2 k_m$ bits are required to describe the quantized gradient vector. The PS recovers all the quantized gradient vectors by performing optimal decoding over the MAC. Thus, for reliable decoding, the transmission rates of the users, i.e., $r_m = d\log_2 k_m$ must be within the MAC capacity region. 

The PS then aggregates the quantized gradients as
\begin{align}
\hat{\mathbf{g}}_t=\frac{1}{M} \sum\limits_{m=1}^{M} \mathbf{Q}(\mathbf{g}_t^{(m)}),
\end{align}
and updates the model using,
\begin{align}
\mathbf{w}_{t+1} = \mathbf{w}_t - \eta_t \hat{\mathbf{g}}_t,
\end{align}
where $\eta_t$ is the learning rate. The updated model is then transmitted back to users for subsequent iterations. 

Suppose that in the $t$th iteration, the dynamic range of the gradient vector of user $m$ is $\Delta_{t,m}$, and the number of quantization levels used is $k_{t,m}$. 
Then, it can be readily checked that $Q(g_{t,i}^{(m)})$ is an unbiased estimator of $g_{t,i}^{(m)}$, i.e., $E \left[ Q(g_{t,i}^{(m)}) \right] = g_{t,i}^{(m)}$. The variance can be computed as, $\text{Var}(Q(g_{t,i}^{(m)})) \leq \Delta_{t,m}^2/4(k_{t,m}-1)^2$.
Therefore, the variance of the quantized gradient vector at user $m$ in iteration $t$ can be bounded as
\begin{align}
\text{Var}(\mathbf{Q}(\mathbf{g}_t^{(m)}))=\sum\limits_{i=1}^{d}\text{Var}(Q(g_{t,i}^{(m)}))\leq \frac{d\Delta_{t,m}^2}{4(k_{t,m}-1)^2}. \label{eq: variance}
\end{align}
We next present our first result which shows how the convergence of the above algorithm depends on the parameters of multi-level stochastic quantization at the users. 


\begin{theorem}
If the loss function $\ell(\cdot)$ is $\lambda$-strongly convex and $\mu$-smooth, with $L$-Lipschitz gradients, then by using a time varying learning rate of $\eta_t = 1/(\lambda t)$, we have the following convergence result:
\begin{align}
     E\left[ \ell(\mathbf{w}_T)\right] - \ell(\mathbf{w}^*) \leq \frac{2\mu}{\lambda^2T^2}\sum\limits_{t=1}^{T} \left(\frac{1}{M^2} \sum\limits_{m=1}^{M} \frac{d\Delta_{t,m}^2}{4(k_{t,m}-1)^2} + L^2\right)\label{eq:convergence}
\end{align}\label{theorem:convergence}
\end{theorem}
The proof of this Theorem is presented in Appendix I. 

\newpage
From Theorem \ref{theorem:convergence}, we observe that the convergence rate depends directly on the following factors: $a)$ the dynamic range of the gradients $\left(\{\Delta_{t,m}\}\right)$ computed by the users, and $b)$ the quantization levels assigned to the users in each iteration. The traditional approach is to assign equal quantization levels to all users, i.e., $k_{t,m} = k$, for all $m,t$. However, the above expression shows that in order to maximize the rate of convergence, users whose gradients have a higher dynamic range must be assigned a higher quantization budget.  On the other hand, if the users are communicating to the PS in a communication constrained setting, such as a MAC, then the quantization budget $k_{t,m}$, which is directly related to the transmission rate cannot exceed the constraints imposed by the capacity region of the MAC.


\subsection{MAC Aware Gradient Quantization} 
Motivated by the above discussion, we propose MAC aware gradient quantization which works as follows. In each iteration $t$, $a)$ users compute their local gradients $\mathbf{g}^{(m)}_{t}$, and describe ${g}^{(m)}_{t, \text{min}}, {g}^{(m)}_{t, \text{max}}$ to the PS. $b)$ using these scalars, PS computes the  dynamic range(s) $\left(\{\Delta_{t,m}={g}^{(m)}_{t, \text{max}} - {g}^{(m)}_{t, \text{min}} \}\right)$ of the gradients for all the users and performs the optimization described in Theorem \ref{theorem:MACaware}. Subsequently, the PS assigns individual quantization budgets (transmission rates) to each user; $c)$ users subsequently quantize their gradients and transmit over the MAC. In the following Theorem, we present the optimization problem using which we can determine the optimal $k_{t,m}^*$'s that maximize the convergence rate. 
\begin{theorem}\label{theorem:MACaware}
At each iteration $t$, the optimal $k_{t,m}^*$'s that give the best convergence rate can be found by solving the following optimization problem,
\begin{align}
\min_{ \{k_{t,m}\}_{m=1}^{M}}& ~\sum\limits_{m=1}^{M} \frac{d\Delta_{t,m}^2}{4(k_{t,m}-1)^2}\\
\text{s.t.}& ~ \sum\limits_{m\in\mathcal{M}}r_{t,m} \leq sC_\mathcal{M},~\mathcal{M}\subset [M],~|\mathcal{M}|=1,\dots,M,\\
& k_{t,m}\in \mathbb{Z}^+,~\forall m
\end{align}
where $r_{t,m}=d\log_2 k_{t,m}$ denotes the transmission rate of user $m$ and $C_{\mathcal{M}}$ denotes the sum capacity of the users in subset $\mathcal{M}$, i.e., $C_{\mathcal{M}} = 0.5\log(1+ \sum_{m\in \mathcal{M}} P_m/\sigma^{2})$, where $\sigma^{2}$ denotes variance of the channel noise. 
\end{theorem}

The above optimization problem falls into the category of constrained integer programming since $k_{t,m}$'s take non-negative integer values. In general, integer programming is considered to be NP-hard problem \cite{schrijver1998theory}. However, one could obtain sub-optimal solutions by relaxing the constraint on $k_{t,m}$'s. For instance, by allowing $k_{t,m}$'s to be real numbers greater or equal to $2$ (so that each user gets at least $1$ bit), it is easy to verify that the above problem becomes a convex optimization problem. One could then either use convex solvers or solve the convex problem analytically by checking KKT conditions, and round the results.
We next show an example for $2$ users, and solve the convex relaxation analytically to gain insights on how the dynamic ranges of the gradients, and the capacity region of MAC impact the resulting quantization budgets.

\subsection{Solution for the Relaxed Optimization Problem with $M=2$}

For $M=2$ users, the relaxed optimization problem ($\mathcal{P}$) is given as follows:
\vspace{-5pt}
\begin{align}
\mathcal{P:} ~ \min_{(k_1, k_2)}& ~\frac{d\Delta_1^2}{4(k_1-1)^2}+\frac{d\Delta_2^2}{4(k_2-1)^2}\\
\text{s.t.}& ~ \begin{aligned}[t] &d\log_2 k_1 \leq sC_1, ~~~ d\log_2 k_2 \leq sC_2 \nonumber \\
&d(\log_2 k_1 + \log_2 k_2) \leq sC_{1,2}
\end{aligned}
\end{align}
The three constraints on rates can be rearranged as follows:
\vspace{-5pt}
\begin{align}
k_1 \leq 2^{\widetilde{C}_1}, ~~ k_2 \leq 2^{\widetilde{C}_2}, ~~ k_1k_2 \leq 2^{\widetilde{C}_{12}},
\end{align}
where $\widetilde{C}_m=sC_m/d,~m=1,2$ and $\widetilde{C}_{12}=sC_{1,2}/d$. As mentioned earlier, the objective function being minimized is a convex function when $k_1$ and $k_2$ are both greater or equal to $2$. The $M=2$-user case can be solved analytically by first forming the following Lagrangian function,
\vspace{-5pt}
\begin{align}
J= \frac{d\Delta_1^2}{4(k_1-1)^2}+\frac{d\Delta_2^2}{4(k_2-1)^2} + \lambda_1 (k_1 - 2^{\widetilde{C}_1})+ \lambda_2 (k_2 - 2^{\widetilde{C}_2}) + \lambda_3 (k_1k_2 - 2^{\widetilde{C}_{1,2}}).
\end{align}
We note that to fully utilize the channel, the sum-rate constraint in $\mathcal{P}$ should be satisfied with equality, i.e., $d(\log_2 k_1 + \log_2 k_2) = sC_{1,2}$ or equivalently, $k_1k_2=2^{\widetilde{C}_{12}}$. By taking the partial derivatives of $J$ with respect to $k_1$ and $k_2$ and checking the KKT conditions, we obtain,
\vspace{-5pt}
\begin{align}
\lambda_1=\lambda_2=0,~\lambda_3 = \frac{d\Delta_1^2}{2k_2(k_1-1)^3} = \frac{d\Delta_2^2}{2k_1(k_2-1)^3}.\label{eq:partialDeriv} 
\end{align}
Using \eqref{eq:partialDeriv} and the sum-rate constraint, i.e., $k_1k_2=2^{\widetilde{C}_{12}}$, we can solve for the optimal quantization budgets. 
\begin{theorem}
For a $2$-user Gaussian MAC, the optimal quantization budgets $k_1^*$ and $k_2^*$ for $\mathcal{P}$ can be found by solving
\vspace{-10pt}
\begin{align}
    \frac{\Delta_1}{\Delta_2} = \left( \frac{2^{\widetilde{C}_{12}} k_1^*(k_1^*-1)^3}{(2^{\widetilde{C}_{12}}-k_1^*)^3} \right)^{1/2},
\end{align}
and subsequently $k_2^*=2^{\widetilde{C}_{1,2}}/k_1^*$, where $\Delta_1$ and $\Delta_2$ are dynamic ranges of gradients at users $1$ and $2$. 
\end{theorem}

\begin{figure}[t]
\centering
	\includegraphics[width=0.5\linewidth]{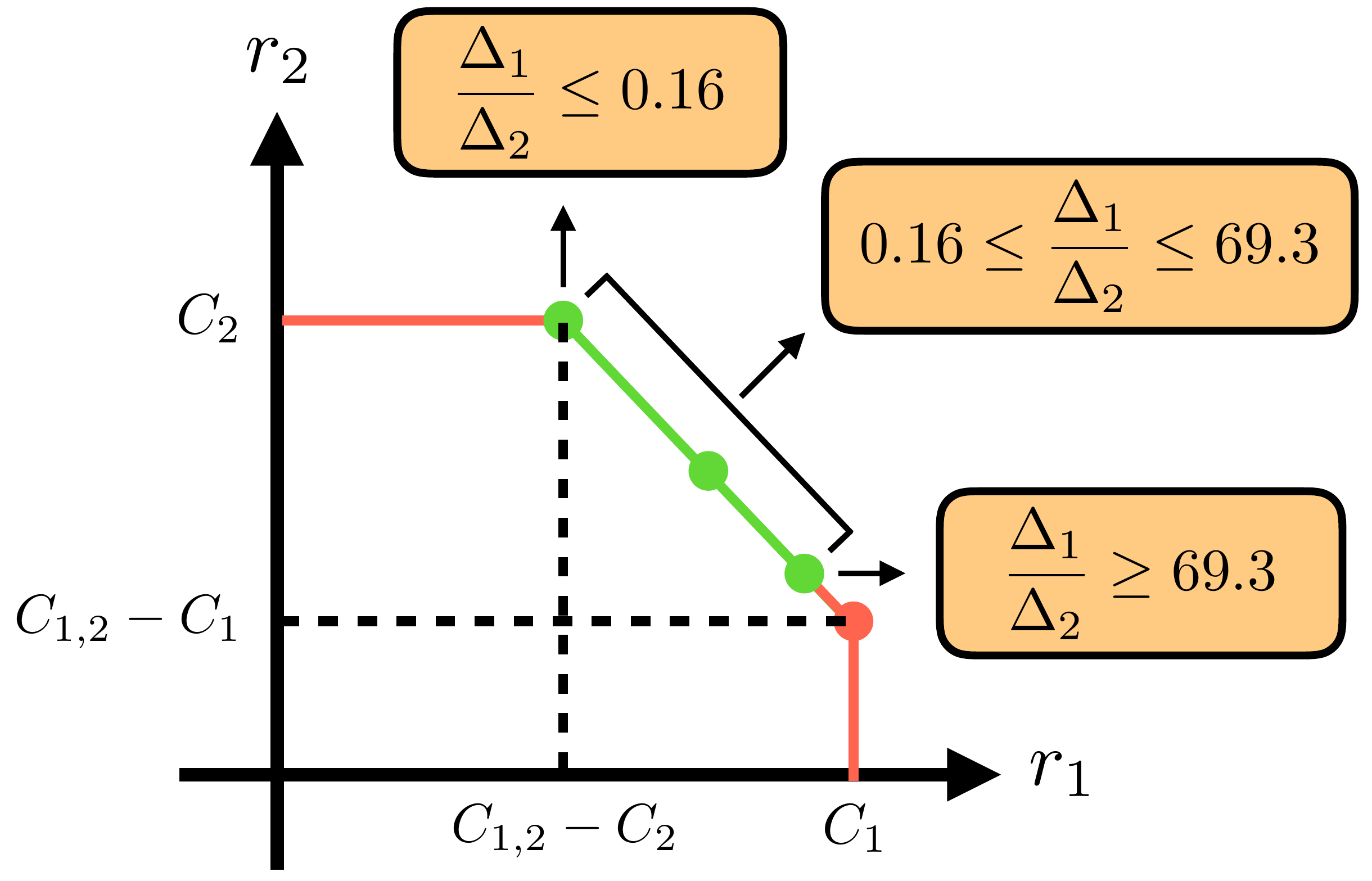}
	\caption{The capacity region of the Gaussian MAC when $P_1=80,P_2=20$. Green area denote points that achieve maximum sum rate.
	\label{Fig:GMAC}}
	\vspace{-10pt}
\end{figure}

We solve $k_1^*$ and $k_2^*$ numerically with the following parameters: we let $d=7850,~s=2d$,  $P_1=80, P_2=20$, so that the individual and sum capacities for this setting are $C_1=3.1699,~C_2=2.1962$ and $C_{1,2}=3.3291$. These lead to $k_1\leq 80.9,~k_2\leq 21$ and $k_1k_2\leq 100.9$. We fix $\Delta_2=50$ and vary $\Delta_1$ from $1$ to $3500$ to understand the impact of the ratio of dynamic range $\Delta_1/\Delta_2$ on the quantization budgets. It can be seen in Fig. \ref{Fig:GMAC} and Table \ref{Fig:table} that by using proposed MAC aware scheme, the PS allocates more rate towards the user whose gradients are more informative (higher dynamic range). For instance, when $\Delta_1/\Delta_2=1$, gradients from both users are equally informative, and both users are assigned equal quantization budgets $k_1=k_2=10$. On one extreme, when  $\Delta_1/\Delta_2\leq 0.16$, gradients from user $2$ are considered  more useful than user $1$, the optimal allocation is  $k_1=4$, $k_2=21$.  On the other extreme, if $\Delta_1/\Delta_2\geq 69.28$, gradients from user $1$ are more informative, hence we see that $k_1=50$, and $k_2=2$. 

\begin{table}[t]
\centering
	\includegraphics[width=0.6\linewidth]{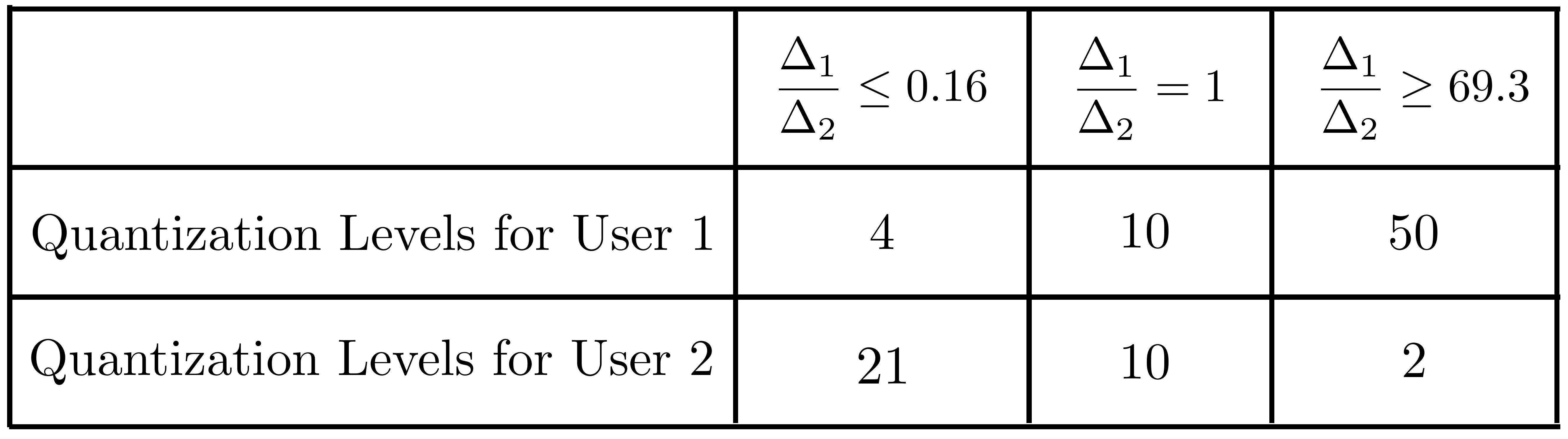}
	\caption{Per-user quantization budget based on ratio of dynamic range of the gradients, i.e.,  $\Delta_1/\Delta_2$ and the capacity region of MAC.
	\label{Fig:table}}
	\vspace{-10pt}
\end{table}


\section{Experiments}
To show the performance of our proposed scheme, we consider MNIST image classification task using single layer neural networks trained on $60000$ training and $10000$ testing samples with $M=2$ users, and a cross-entropy loss function. The dimensionality of the classifier model is $d=7850$. We assume that user $1$'s data set $\mathcal{D}_1$ consists of images  belonging to digits '0' and '1', whereas the data set of user $2$ consists of all the $10$ digits. The channel noise variance is set as $\sigma^2=1$, and the total transmit power per iteration is set as $\bar{P}=100$. We use the MAC for $s=2d$ channel uses for each iteration. 


\begin{figure}[t]
\centering
	\includegraphics[width=0.47\linewidth]{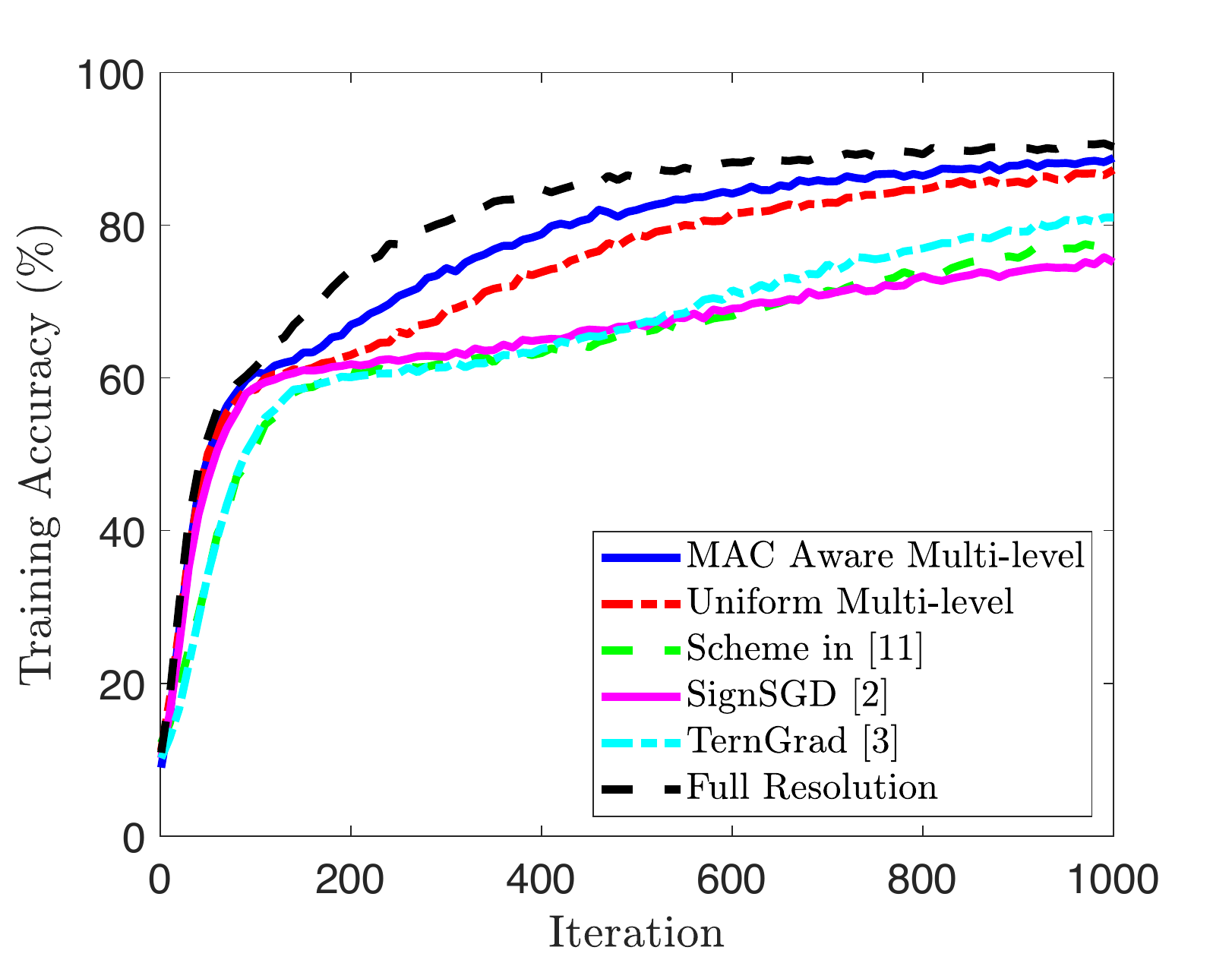}
	\caption{Training accuracy comparison between MAC aware gradient quantization, uniform rate allocation, digital scheme proposed in \cite{AmiDen2019}, SignSGD \cite{signSGD2018}, TernGrad \cite{TernGrad2017}, and full resolution when the total transmit power per iteration is $\bar{P}=100$ and $s=2d$.
	\label{Fig:2usersAcc}}
	\vspace{-10pt}
\end{figure}
In Fig. \ref{Fig:2usersAcc}, we let $P_1=0.95\bar{P}$ and $P_2=0.05\bar{P}$, and compare the proposed MAC aware gradient quantization scheme with the following schemes: $a)$ uniform rate allocation subject to MAC capacity constraints, $b)$ a recently proposed digital scheme in \cite{AmiDen2019},  $c)$ SignSGD, which uses $1$ bit quantization per dimension for each user \cite{signSGD2018}, and $d)$ TernGrad    \cite{TernGrad2017}, which uses three levels $\{-1, 0, +1\}$ to quantize each dimension of the gradient. We also plot the non-quantized full resolution scheme as a baseline. In the digital scheme proposed in \cite{AmiDen2019}, all but the highest $q_t$ and lowest $q_t$ gradient values are set to zero. The remaining gradient values are then split into two groups depending on their signs. The mean of elements in each group is computed, denoted by $\alpha_{\text{avg}}^+$ and $\alpha_{\text{avg}}^-$. If $\alpha_{\text{avg}}^+ >  |\alpha_{\text{avg}}^-|$ ($\alpha_{\text{avg}}^+ <  |\alpha_{\text{avg}}^-|$), all remaining positive (negative) values will be set to $\alpha_{\text{avg}}^+$ ($\alpha_{\text{avg}}^-$). Each user then transmits the location of $q_t$ non-zero values and a scalar (using $c$ bits) to describe the average value at each iteration. Therefore, the communication cost is $\log_2\binom{d}{q_t}+c$. This scheme \cite{AmiDen2019} is fundamentally different than the one proposed in this paper, and, moreover, the quantization budget $q_t$ is the same for all users. As shown in Fig. \ref{Fig:2usersAcc}, the proposed MAC aware multi-level scheme outperforms the uniform multilevel scheme, the scheme in \cite{AmiDen2019}, SignSGD and TernGrad. This is due to the fact that $\log\binom{d}{q_t}$ grows exponentially as $q_t$ increases. In addition, the rates are limited by the user with the worst channel. Therefore, as it reaches the capacity of the user with the worst channel, $q_t$ is still small compared to $d$. Other schemes such as SignSGD and TernGrad suffer from underutilization of channel resources, as they use a fixed quantization budget ($1$ bit, and $2$ bits respectively per gradient dimension). We also show the testing accuracy of each scheme at the end of $1000$ iterations (see Table \ref{Fig:testAcc}). They are consistent with Fig. \ref{Fig:2usersAcc} where our proposed scheme is the closest to full resolution.
\begin{table}[t]
\centering
	\includegraphics[width=0.6\linewidth]{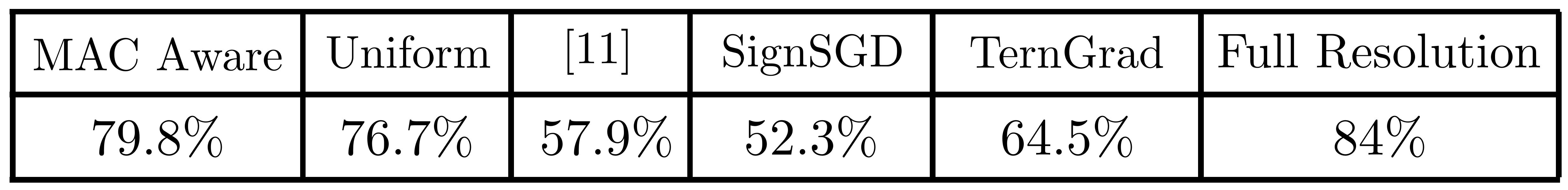}
	\caption{Comparison of test accuracy after $T=1000$ iterations.
	\label{Fig:testAcc}}
	\vspace{-15pt}
\end{table}
\begin{figure}[t]
\centering
	\includegraphics[width=0.47\linewidth]{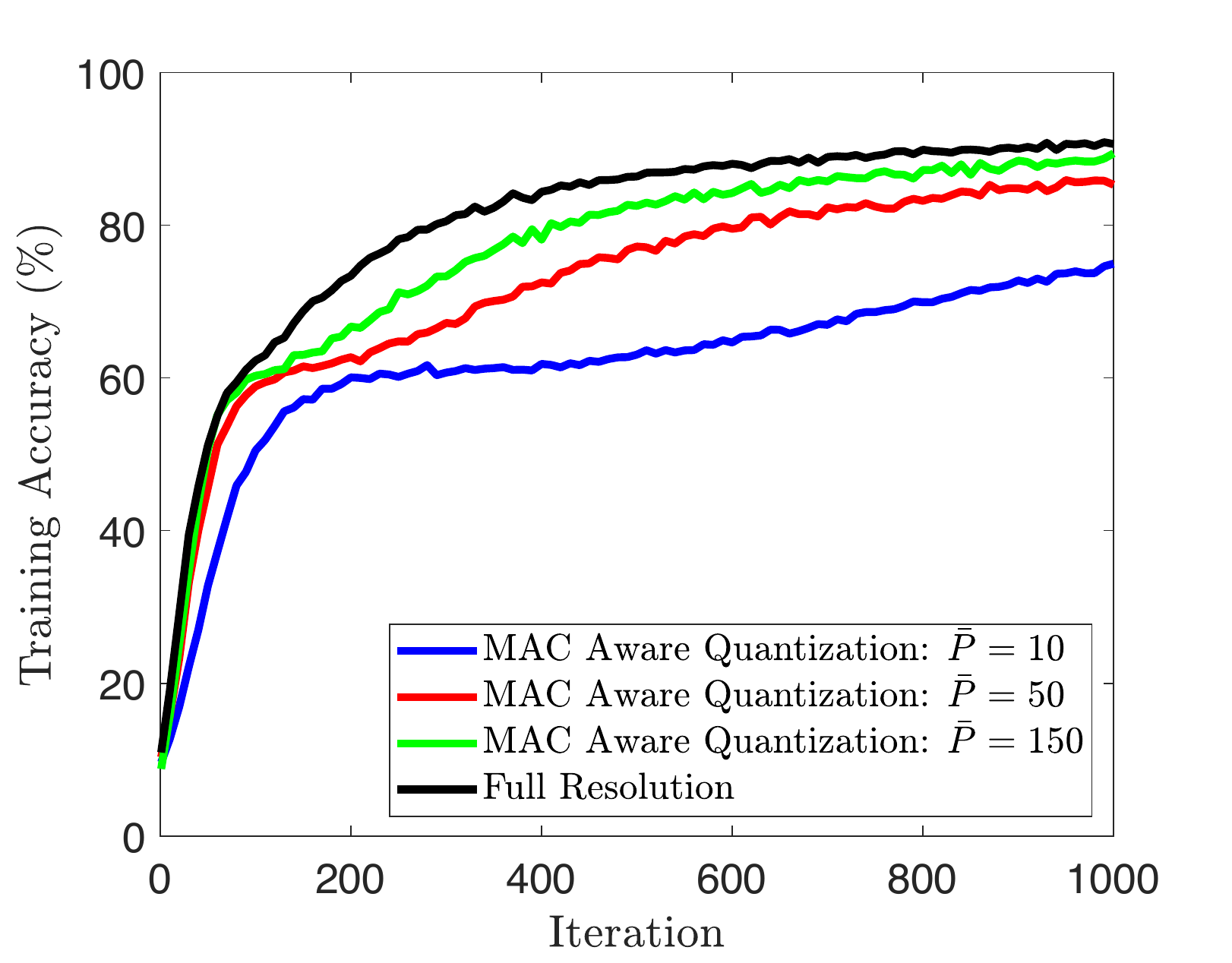}
	\vspace{-5pt}
	\caption{Training accuracy comparison for MAC aware gradient quantization with total power per iteration $\bar{P}=10,50,150$, and $P_1=0.8\bar{P}$ and $P_2=0.2\bar{P}$.
	\label{Fig:variousPower}}
	\vspace{-10pt}
\end{figure}
\begin{table}[t]
\centering
	\includegraphics[width=0.5\linewidth]{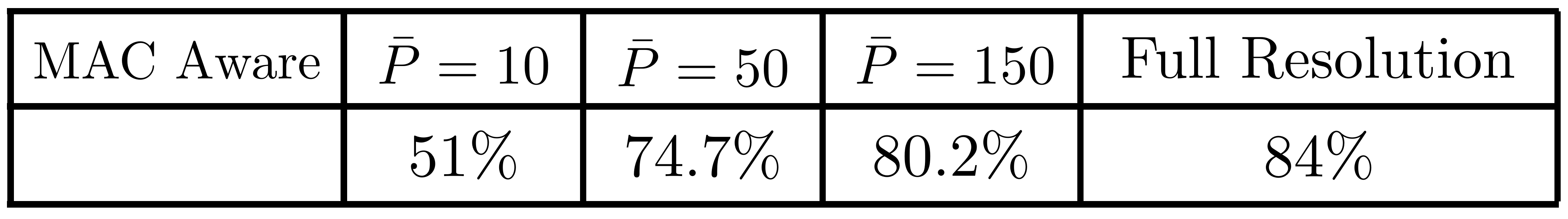}
	\vspace{-5pt}
	\caption{Test accuracy for proposed scheme as a function of total power.
	\label{Fig:testAccVarPower}}
	\vspace{-10pt}
\end{table}

For Fig. \ref{Fig:variousPower}, we set $s=1.5d$, $P_1=0.8\bar{P}$ and $P_2=0.2\bar{P}$, and vary $\bar{P}$ to see the impact of increasing power, and thus, a larger capacity region. It can be seen in Fig. \ref{Fig:variousPower} that the performance improves monotonically with the increase in total power. The testing accuracy at the end of $T=1000$ iterations is shown in Table \ref{Fig:testAccVarPower} as a function of the total power. 

\section{Conclusions \label{Sec:Conclusion}}
In this paper, we considered the problem of MAC aware gradient quantization for federated learning. We showed that when designing digital FL schemes over MACs, there are new opportunities to assign different amount of resources (such as quantization rates) to different users based on a) the informativeness of the gradients at each user, captured by their dynamic range, and b) the underlying channel conditions. We  studied and analyzed a \textit{channel aware quantization} scheme and showed that it outperforms uniform quantization and other existing digital schemes.  
An interesting future direction is to explore if other quantization schemes (for instance, the scheme in \cite{AmiDen2019}, or gradient sparsification schemes in \cite{sparse-1, sparse-2}) can be optimized (with limited interaction with the PS) as a function of the underlying communication channel such as MAC. 

\section*{Appendix I: Proof of Theorem \ref{theorem:convergence}}

Standard convergence results in \cite{RakSha2012} have shown that for a loss function $\ell(.)$ that is $\lambda$-strongly convex and $\mu$-smooth w.r.t. $\mathbf{w}^*$, using SGD with stochastic unbiased gradients, bounded second order moments, i.e., $E[\Vert \hat{\mathbf{g}}_t \Vert_2^2]\leq G^2$, with a  learning rate of $\eta_t=1/\lambda t$ can achieve a convergence result: 
\vspace{-5pt}
\begin{align}
    E\left[ \ell(\mathbf{w}_T)\right] - \ell(\mathbf{w}^*)\leq \frac{2\mu G^2}{\lambda^2T}.\label{eq:convBound}
\end{align}
There are two distinctions between our bound and \eqref{eq:convBound}. First, the randomness in our scheme comes from quantizing the gradients instead of randomly selecting data points. Second, as users can have different quantization budgets per iteration, the resulting  variance is iteration dependent, i.e., $E[\Vert \hat{\mathbf{g}}_t \Vert_2^2]\leq G_t^2$. By slightly modifying the proof  in \cite{RakSha2012}, it is possible to prove the following convergence result (proof omitted due to space):
\vspace{-10pt}
\begin{align}
    E\left[ \ell(\mathbf{w}_T)\right] - \ell(\mathbf{w}^*) \leq \frac{2\mu }{\lambda^2T}\left(\sum\limits_{t=1}^{T}G_t^2/T\right).\label{eq:convBound-Our}
\end{align}
Theorem \ref{theorem:convergence} now follows directly by plugging in the values of $G_{t}^{2}$, which can be computed as:
\vspace{-5pt}
\begin{align}
    E[\Vert \hat{\mathbf{g}}_t \Vert_2^2] &= \text{Var}(\hat{\mathbf{g}}_t) + \Vert \mathbf{g}_t \Vert_2^2\nonumber\\
    &= \frac{1}{M^2}\sum\limits_{m=1}^{M} \text{Var}(Q(\mathbf{g}_t^{(m)})) + \Vert \mathbf{g}_t \Vert_2^2\nonumber\\ &\overset{(a)}\leq \frac{1}{M^2} \sum\limits_{m=1}^{M} \frac{d\Delta_{t,m}^2}{4(k_{t,m}-1)^2} + L^2 \triangleq G_t^2,
\end{align}
where (a) follows from  \eqref{eq: variance} and Lipschitz assumption, i.e.,  $\Vert \mathbf{g}_t \Vert_2^2\leq L^2$.

\bibliographystyle{IEEEtran}
\bibliography{Ref}

\begin{thebibliography}{10}
\providecommand{\url}[1]{#1}
\csname url@samestyle\endcsname
\providecommand{\newblock}{\relax}
\providecommand{\bibinfo}[2]{#2}
\providecommand{\BIBentrySTDinterwordspacing}{\spaceskip=0pt\relax}
\providecommand{\BIBentryALTinterwordstretchfactor}{4}
\providecommand{\BIBentryALTinterwordspacing}{\spaceskip=\fontdimen2\font plus
\BIBentryALTinterwordstretchfactor\fontdimen3\font minus
  \fontdimen4\font\relax}
\providecommand{\BIBforeignlanguage}[2]{{%
\expandafter\ifx\csname l@#1\endcsname\relax
\typeout{** WARNING: IEEEtran.bst: No hyphenation pattern has been}%
\typeout{** loaded for the language `#1'. Using the pattern for}%
\typeout{** the default language instead.}%
\else
\language=\csname l@#1\endcsname
\fi
#2}}
\providecommand{\BIBdecl}{\relax}
\BIBdecl

\bibitem{1BitSGD}
F.~Seide, H.~Fu, J.~Droppo, G.~Li, and D.~Yu, ``{1-Bit Stochastic Gradient
  Descent and Application to Data-Parallel Distributed Training of Speech
  DNNs},'' in \emph{Interspeech 2014}, September 2014.

\bibitem{signSGD2018}
J.~Bernstein, Y.-X. Wang, K.~Azizzadenesheli, and A.~Anandkumar, ``sign{SGD}:
  Compressed optimisation for non-convex problems,'' in \emph{Proceedings of
  the 35th International Conference on Machine Learning}, ser. Proceedings of
  Machine Learning Research, vol.~80, 10--15 Jul 2018, pp. 560--569.

\bibitem{TernGrad2017}
W.~Wen, C.~Xu, F.~Yan, C.~Wu, Y.~Wang, Y.~Chen, and H.~Li, ``{TernGrad: Ternary
  Gradients to Reduce Communication in Distributed Deep Learning},'' in
  \emph{Advances in Neural Information Processing Systems 30}, 2017, pp.
  1509--1519.

\bibitem{sparse-1}
\BIBentryALTinterwordspacing
A.~F. Aji and K.~Heafield, ``Sparse communication for distributed gradient
  descent,'' \emph{CoRR}, vol. abs/1704.05021, 2017. [Online]. Available:
  \url{http://arxiv.org/abs/1704.05021}
\BIBentrySTDinterwordspacing

\bibitem{sparse-2}
J.~Wangni, J.~Wang, J.~Liu, and T.~Zhang, ``Gradient sparsification for
  communication-efficient distributed optimization,'' in \emph{Advances in
  Neural Information Processing Systems 31}, 2018, pp. 1299--1309.

\bibitem{sparse-3}
\BIBentryALTinterwordspacing
N.~Dryden, S.~A. Jacobs, T.~Moon, and B.~Van~Essen, ``Communication
  quantization for data-parallel training of deep neural networks,'' in
  \emph{Proceedings of the Workshop on Machine Learning in High Performance
  Computing Environments}, ser. MLHPC '16.\hskip 1em plus 0.5em minus
  0.4em\relax Piscataway, NJ, USA: IEEE Press, 2016, pp. 1--8. [Online].
  Available: \url{https://doi.org/10.1109/MLHPC.2016.4}
\BIBentrySTDinterwordspacing

\bibitem{sparse-4}
\BIBentryALTinterwordspacing
Y.~Lin, S.~Han, H.~Mao, Y.~Wang, and W.~J. Dally, ``Deep gradient compression:
  Reducing the communication bandwidth for distributed training,'' \emph{CoRR},
  vol. abs/1712.01887, 2017. [Online]. Available:
  \url{http://arxiv.org/abs/1712.01887}
\BIBentrySTDinterwordspacing

\bibitem{sparse-5}
\BIBentryALTinterwordspacing
F.~Sattler, S.~Wiedemann, K.~M{\"{u}}ller, and W.~Samek, ``Sparse binary
  compression: Towards distributed deep learning with minimal communication,''
  \emph{CoRR}, vol. abs/1805.08768, 2018. [Online]. Available:
  \url{http://arxiv.org/abs/1805.08768}
\BIBentrySTDinterwordspacing

\bibitem{QSGD}
D.~Alistarh, D.~Grubic, J.~Li, R.~Tomioka, and M.~Vojnovic, ``{QSGD:
  Communication-Efficient SGD via Gradient Quantization and Encoding},'' in
  \emph{Advances in Neural Information Processing Systems 30}, 2017, pp.
  1709--1720.

\bibitem{klevel}
A.~T. Suresh, F.~X. Yu, S.~Kumar, and H.~B. McMahan, ``Distributed mean
  estimation with limited communication,'' in \emph{Proceedings of the 34th
  International Conference on Machine Learning}, 2017, p. 3329–3337.

\bibitem{AmiDen2019}
\BIBentryALTinterwordspacing
M.~M. Amiri and D.~G{\"{u}}nd{\"{u}}z, ``Machine learning at the wireless edge:
  Distributed stochastic gradient descent over-the-air,'' \emph{CoRR}, vol.
  abs/1901.00844, 2019. [Online]. Available:
  \url{http://arxiv.org/abs/1901.00844}
\BIBentrySTDinterwordspacing

\bibitem{AmiDen2019-2}
M.~M. {Amiri} and D.~G{\"{u}}nd{\"{u}}z, ``Over-the-air machine learning at the
  wireless edge,'' in \emph{2019 IEEE 20th International Workshop on Signal
  Processing Advances in Wireless Communications (SPAWC)}, July 2019, pp. 1--5.

\bibitem{AmiDen2019-3}
\BIBentryALTinterwordspacing
M.~M. Amiri and D.~G{\"{u}}nd{\"{u}}z, ``Federated learning over wireless
  fading channels,'' \emph{CoRR}, vol. abs/1907.09769, 2019. [Online].
  Available: \url{http://arxiv.org/abs/1907.09769}
\BIBentrySTDinterwordspacing

\bibitem{AmiDum2019}
\BIBentryALTinterwordspacing
M.~M. Amiri, T.~M. Duman, and D.~G{\"{u}}nd{\"{u}}z, ``Collaborative machine
  learning at the wireless edge with blind transmitters,'' \emph{CoRR}, vol.
  abs/1907.03909, 2019. [Online]. Available:
  \url{http://arxiv.org/abs/1907.03909}
\BIBentrySTDinterwordspacing

\bibitem{AbaOzf2019}
\BIBentryALTinterwordspacing
M.~S.~H. Abad, E.~Ozfatura, D.~G{\"{u}}nd{\"{u}}z, and O.~Ercetin,
  ``Hierarchical federated learning across heterogeneous cellular networks,''
  \emph{CoRR}, vol. abs/1909.02362, 2019. [Online]. Available:
  \url{http://arxiv.org/abs/1909.02362}
\BIBentrySTDinterwordspacing

\bibitem{CheYan2019}
\BIBentryALTinterwordspacing
M.~Chen, Z.~Yang, W.~Saad, C.~Yin, H.~V. Poor, and S.~Cui, ``A joint learning
  and communications framework for federated learning over wireless networks,''
  \emph{CoRR}, vol. abs/1909.07972, 2019. [Online]. Available:
  \url{http://arxiv.org/abs/1909.07972}
\BIBentrySTDinterwordspacing

\bibitem{YanJia2018}
\BIBentryALTinterwordspacing
K.~Yang, T.~Jiang, Y.~Shi, and Z.~Ding, ``Federated learning via over-the-air
  computation,'' \emph{CoRR}, vol. abs/1812.11750, 2018. [Online]. Available:
  \url{http://arxiv.org/abs/1812.11750}
\BIBentrySTDinterwordspacing

\bibitem{ZenDu2019}
\BIBentryALTinterwordspacing
Q.~Zeng, Y.~Du, K.~K. Leung, and K.~Huang, ``Energy-efficient radio resource
  allocation for federated edge learning,'' \emph{CoRR}, vol. abs/1907.06040,
  2019. [Online]. Available: \url{http://arxiv.org/abs/1907.06040}
\BIBentrySTDinterwordspacing

\bibitem{ZhuWan2020}
G.~{Zhu}, Y.~{Wang}, and K.~{Huang}, ``Broadband analog aggregation for
  low-latency federated edge learning,'' \emph{IEEE Transactions on Wireless
  Communications}, vol.~19, no.~1, pp. 491--506, Jan. 2020.

\bibitem{SunZho2019}
\BIBentryALTinterwordspacing
Y.~Sun, S.~Zhou, and D.~Gündüz, ``Energy-aware analog aggregation for
  federated learning with redundant data,'' \emph{CoRR}, vol. abs/1911.00188,
  2019. [Online]. Available: \url{http://arxiv.org/abs/1911.00188}
\BIBentrySTDinterwordspacing

\bibitem{WanTuo2018}
\BIBentryALTinterwordspacing
S.~Wang, T.~Tuor, T.~Salonidis, K.~K. Leung, C.~Makaya, T.~He, and K.~Chan,
  ``When edge meets learning: Adaptive control for resource-constrained
  distributed machine learning,'' \emph{CoRR}, vol. abs/1804.05271, 2018.
  [Online]. Available: \url{http://arxiv.org/abs/1804.05271}
\BIBentrySTDinterwordspacing

\bibitem{SerCoh2019}
\BIBentryALTinterwordspacing
T.~Sery and K.~Cohen, ``On analog gradient descent learning over multiple
  access fading channels,'' \emph{CoRR}, vol. abs/1908.07463, 2019. [Online].
  Available: \url{http://arxiv.org/abs/1908.07463}
\BIBentrySTDinterwordspacing

\bibitem{SerCoh2019-2}
T.~{Sery} and K.~{Cohen}, ``A sequential gradient-based multiple access for
  distributed learning over fading channels,'' in \emph{2019 57th Annual
  Allerton Conference on Communication, Control, and Computing (Allerton)},
  Sep. 2019, pp. 303--307.

\bibitem{Cover-thomas}
T.~M. Cover and J.~A. Thomas, \emph{Elements of Information Theory}.\hskip 1em
  plus 0.5em minus 0.4em\relax Wiley-Interscience, 2006.

\bibitem{AgaSur2018}
\BIBentryALTinterwordspacing
N.~Agarwal, A.~T. Suresh, F.~Yu, S.~Kumar, and H.~B. Mcmahan, ``{cpSGD:
  Communication-efficient and differentially-private distributed SGD},''
  \emph{CoRR}, vol. abs/1805.10559, 2018. [Online]. Available:
  \url{http://arxiv.org/abs/1805.10559}
\BIBentrySTDinterwordspacing

\bibitem{schrijver1998theory}
A.~Schrijver, \emph{Theory of linear and integer programming}.\hskip 1em plus
  0.5em minus 0.4em\relax John Wiley \& Sons, 1998.

\bibitem{RakSha2012}
\BIBentryALTinterwordspacing
A.~Rakhlin, O.~Shamir, and K.~Sridharan, ``Making gradient descent optimal for
  strongly convex stochastic optimization,'' \emph{CoRR}, vol. abs/1109.5647,
  2012. [Online]. Available: \url{http://arxiv.org/abs/1109.5647}
\BIBentrySTDinterwordspacing

\end{thebibliography}

\end{document}